\documentstyle[12pt,graphicx,caption2]{article}

\begin{document}

\title{On the $n\bar{n}$ transitions in the medium}
\author{V.I.Nazaruk\\
Institute for Nuclear Research of RAS, 60th October\\
Anniversary Prospect 7a, 117312 Moscow, Russia.*}

\date{}
\maketitle
\bigskip

\begin{abstract}
The corrections to the model of $n\bar{n}$ transitions in the medium
and additional baryon-number-violating processes are considered. We
focus on the time-dependence since it is different for the $n\bar{n}$
transition followed by annihilation (basic process) and processes
mentioned above.

\end{abstract}

\vspace{5mm}

\vspace{5mm}

\vspace{1cm}

*E-mail: nazaruk@inr.ru

\newpage
\setcounter{equation}{0}
\section{Introduction}
In the standard calculations of $ab$ oscillations in the medium [1-3] the
interaction of particles $a$ and $b$ with the matter is described by the potentials
$U_{a,b}$. ${\rm Im}U_b$ is responsible for loss of $b$-particle intensity. In
particular, this model is used for the $n\bar{n}$ transitions in a medium [4-7].
In [8,9] it was shown that one-particle model mentioned above does not describe the
total $ab$ transition probability as well as the channel corresponding to absorption
of the $b$-particle.

In a resent paper [10] the field-theoretical approach to particle oscillations in 
absorbing matter has been proposed. It was studied by the example of $n\bar{n}$ 
transitions. In this case the main channels are the annihilation and scattering one. 
The $n\bar{n}$ transition in the medium followed by annihilation
\begin{equation}
(n-\mbox{medium})\rightarrow (\bar{n}-\mbox{medium})\rightarrow M,
\end{equation}
($M$ are the annihilation mesons) is shown in Fig. 1a. The $n\bar{n}$ transition with 
$\bar{n}$ in the final state is shown in Fig. 1b. The block $T^{\bar{n}}_{fi}$ involves 
all the $\bar{n}$-medium interactions followed by annihilation including the antineutron 
rescattering in the initial state; $S=1+iT$. Similarly, $T^{\bar{n}}_{ii}$ involves 
scattering proper and annihilation loops. Due to this the antineutron propagator is bare. 
This model is considered in more detail below.

Let $\sigma _a$ and $\sigma _s$ are the cross sections of free-space $\bar{n}N$
annihilation and $\bar{n}N$ scattering, respectively. At the low energies $\sigma 
_a>2.5\sigma _s$ [11,12]. Hence the annihilation channel (Fig. 1a) is dominant: in 
the first stage of $\bar{n}$-medium interaction the annihilation occurs. At the 
nuclear densities only the diagram 1a is essential because the antineutron annihilates 
in a time $\tau _a\sim 1/\Gamma $, where $\Gamma $ is the annihilation width of 
$\bar{n}$ in the medium. 

In [10] the diagrams 1 have been calculated. The corrections to this model and other 
baryon-number-violating ($\Delta B=2$) processes [13] (see Figs. 2 and 3) can 
essentially change the result. Besides, it turns out that the time-dependence of the 
processes shown in Figs. 1 and 2 is different. This is non-trivial circumstance and 
we focus on this point. The heart of the problem is as follows. The processes shown in 
Fig. 2 are described by the exponential decay law. The diagram 1a contains the 
infrared divergence conditioned by zero momentum transfer in the $n\bar{n}$ transition 
vertex. This is unremovable perculiarity. The fact that amplitude is singular means that 
the standard $S$-matrix approach is inapplicable. As a consequence the other surprises 
take place as well, in particular, the different functional structure of the result and
non-exponential behavior. These problems are studied below. 

\section{Calculation}
First of all we consider the incoherent contribution of the diagrams 2. In Fig. 2a a meson 
is radiated before the $n\bar{n}$ transition. The interaction Hamiltonian has the form
\begin{equation}
{\cal H}_I=g\Psi ^+_n\Phi \Psi _n+{\cal H}_{n\bar{n}}+{\cal H},
\end{equation}
\begin{equation}
{\cal H}_{n\bar{n}}=\epsilon \Psi ^+_{\bar{n}}\Psi _n+{\rm H.c.}
\end{equation}
Here ${\cal H}_{n\bar{n}}$ is the Hamiltonian of the free-space $n\bar{n}$ transition
[6,10], $\epsilon $ is the off-diagonal mass (transition mass); $\epsilon =1/\tau _
{n\bar{n}}$, where $\tau _{n\bar{n}}$ is the free-space $n\bar{n}$ oscillation time;
${\cal H}$ is the Hamiltonian of $\bar{n}$-medium interaction taken in the general
form. In the following the background neutron potential is omitted; $m_n=m_{\bar{n}}=m$. 
The neutron wave function is $n_p(x)=\Omega ^{-1/2}\exp (-ipx)$, were $p=(p_0,{\bf p})$ 
and $p_0=m+{\bf p}^2/2m$.

\begin{figure}[h]
%  \reflectbox{\includegraphics[height=.3\textheight]{golfer}}
  {\includegraphics[height=.3\textheight]{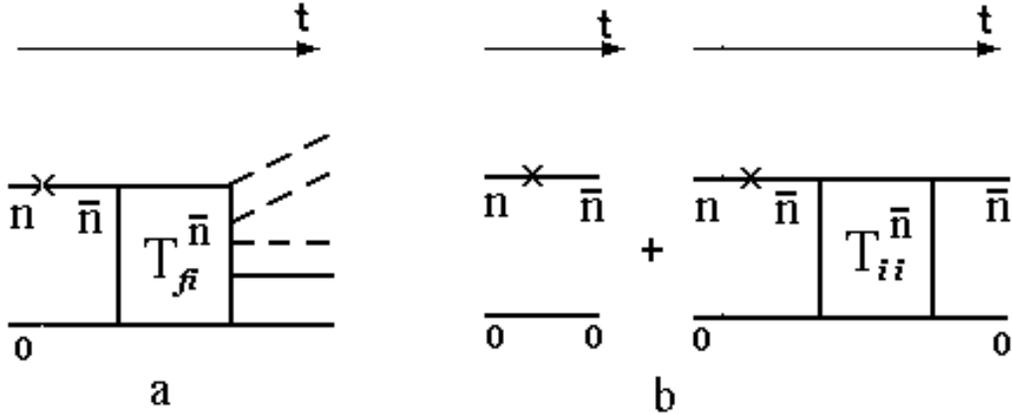}}
  \caption{{\bf a} $n\bar{n}$ transition in the medium followed by annihilation.
{\bf b} $n\bar{n}$ transition in the medium with $\bar{n}$ in the final state.}
\end{figure}

For the process amplitude $M_{2a}$ one obtains
\begin{equation}
M_{2a}=gG\epsilon GM^{(n-1)},
\end{equation}
\begin{equation}
G=\frac{1}{p_0-q_0-m-({\bf p}-{\bf q})^2/2m+i0},
\end{equation}
where $q$ is the 4-momenta of meson radiated, $M^{(n)}$ is the amplitude of antineutron
annihilation in the medium in the $(n)$ mesons. It is defined in the usual manner:
\begin{equation}
iT_{fi}^{\bar{n}}=<\!(n)0\!\mid T\exp (-i\int dx{\cal H}(x))-1\mid\!0\bar{n}_{p-q}\!>=
N(2\pi )^4\delta ^4(p_f-p_i)M^{(n)},
\end{equation}
$S=1+iT$. Here $\mid\!0\bar{n}_{p-q}\!>$ is the state of the medium containing the
$\bar{n}$ with the 4-momentum $p-q$, $<\!(n)\!\mid $ denotes the final state of the
annihilation mesons, $N$ includes the normalization factors of the wave functions.
Since $M^{(n)}$ involves all the $\bar{n}$-medium interactions followed by
annihilation including the antineutron rescattering in the initial state, the
antineutron propagator $G$ is bare; the $\bar{n}$ self-energy $\Sigma =0$.

\begin{figure}[h]
%  \reflectbox{\includegraphics[height=.3\textheight]{golfer}}
  {\includegraphics[height=.3\textheight]{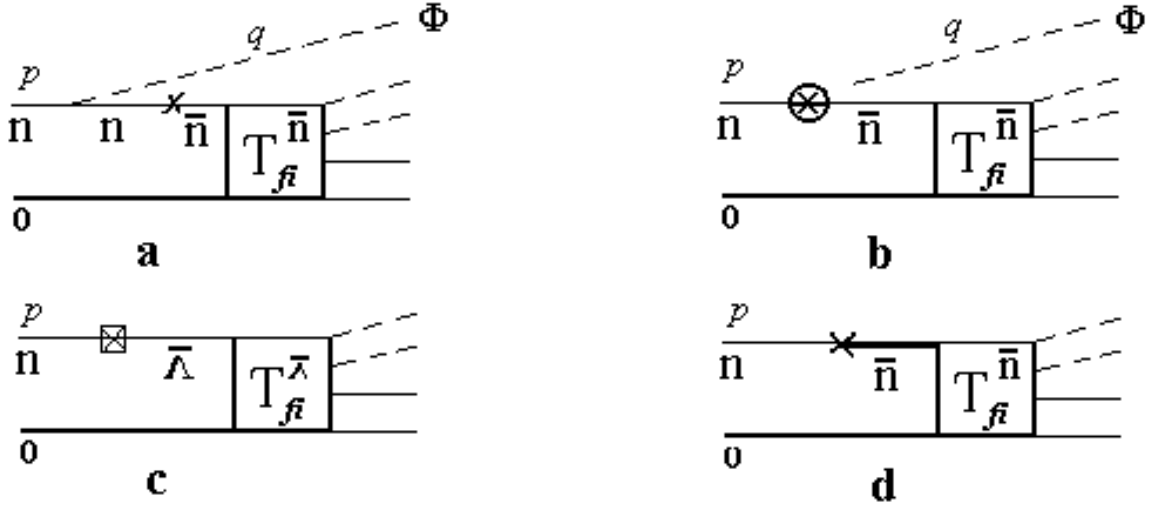}}
  \caption{Corrections to the model shown in Fig. 1a ({\bf a} and {\bf d}) and
additional baryon-number-violating processes ({\bf b} and {\bf c}).}
\end{figure}

Let $\Gamma _{2a}$ and $\Gamma ^{(n)}$ be the widths corresponding to the Fig. 2a
and annihilation width of $\bar{n}$ in the $(n)$ mesons, respectively; $\Gamma =
\sum_{(n)}\Gamma ^{(n)}$. Taking into account that $\Gamma ^{(n)}$ is a smooth
function of $\sqrt{s}$ and summing over $(n)$, it is easy to get the estimation:
\begin{equation}
\Gamma _{2a}\approx 5\cdot 10^{-3}g^2\frac{\epsilon ^2}{m^2_{\Phi }}\Gamma
\approx \frac{\epsilon ^2}{m^2_{\Phi }}\Gamma .
\end{equation}
The time-dependence is determined by the exponential decay law:
\begin{equation}
W_{2a}(t)=1-e^{-\Gamma _{2a}t}\sim \Gamma _{2a}t.
\end{equation}
If $q\rightarrow 0$, the amplitude $M_{2a}$ increases since $G\rightarrow G_s$,
\begin{equation}
G_s=\frac{1}{p_0-m-{\bf p}^2/2m}\sim \frac{1}{0}.
\end{equation}
(The limiting transition $q\rightarrow 0$ for the diagram 2a is an imaginary
procedure because in the vertex $n\rightarrow n\Phi $ the real meson is
escaped and so $q_0\geq m_{\Phi }$.) The fact that the amplitude increases is
essential for us because for Fig. 1a $q=0$.

Consider now the baryon-number-violating decay $n\rightarrow \bar{n}\Phi $ [13]
shown in Fig. 2b. It leads to the same final state, as the processes depicted in
Figs. 1a and 2a. As with ${\cal H}_{n\bar{n}}$, the Hamiltonian of the decay
$n\rightarrow \bar{n}\Phi $ is taken in the scalar form $\epsilon _{\Phi }\Psi 
^+_{\bar{n}}\Phi \Psi _n+$H.c. Then
\begin{equation}
{\cal H}_I=\epsilon _{\Phi }\Psi ^+_{\bar{n}}\Phi \Psi _n+{\cal H},
\end{equation}
$\epsilon _{\Phi }$ is dimensionless. The process amplitude is
\begin{equation}
M_{2b}=\epsilon _{\Phi }GM^{(n-1)}.
\end{equation}
The antineutron propagator $G$ is given by (5). Denoting $\mid {\bf q}\mid=q$, for
the decay width $\Gamma _{2b}$ one obtains
\begin{equation}
\Gamma _{2b}=\frac{\epsilon _{\Phi }^2}{(2\pi )^2}\int dq\frac{q^2}{q_0}G^2\Gamma (q),
\end{equation}
$q_0^2=q^2+m_{\Phi }^2$. As with Fig. 2a, the time-dependence is determined by the 
exponential decay law $W_{2b}(t)\sim \Gamma _{2b}t$.

In Fig. 2c the baryon-number-violating conversion $n\rightarrow \bar{\Lambda }$ in
the medium [13] is shown. The amplitude is
\begin{equation}
M_{2c}=\epsilon _{\Lambda } \frac{1}{p_0-m_{\Lambda }-{\bf p}^2/2m_{\Lambda }+i0}
M^{(n)}_{\Lambda },
\end{equation}
where $\epsilon _{\Lambda }$ corresponds to the vertex $n\rightarrow \bar{\Lambda }$,
$M^{(n)}_{\Lambda }$ is the amplitude of $\bar{\Lambda }$ annihilation in the medium
in the $(n)$ mesons. Contrary to Fig. 1a, there is no infrared singularity because
$m_{\Lambda }\neq m$. We also note that this process cannot produce interference,
since there is $K$-meson in the final state.

Let us suppose that there is residual $\bar{n}$-medium interaction $V$ (scalar field)
which cannot be involved in ${\cal H}$, i.e. the block $T^{\bar{n}}_{fi}$ (see Fig. 2d).
In this case it is included in the in medium antineutron Green function
\begin{equation}
G_m=G_s+G_sVG_s+...=\frac{1}{(1/G_s)-V}=-\frac{1}{V}.
\end{equation}
The process amplitude is $M_{2d}=\epsilon G_mM^{(n)}$. The process width $\Gamma _{2d}$ 
is found to be 
\begin{equation} 
\Gamma _{2d}=\frac{\epsilon ^2}{V^2}\Gamma .  
\end{equation}
In fact, the correct definition of antineutron annihilation amplitude (see (6))
excludes the diagrams like 2d. It was adduced only for the purpose of illustration
since the models with the dressed antineutron propagator were at one time discussed.
It also demonstrates the solution stability to the possible perturbations, which
seems important. 

In Figs. 2c and 2d the vertexes $n\rightarrow \bar{\Lambda }$ and $n\rightarrow
\bar{n}$ are depicted by 2-tail diagram. Nevertheless, the amplitudes $M_{2c}$ and
$M_{2d}$ are non-singular because $m_{\Lambda }\neq m$ (Fig. 2c) and $m_{\bar{n}}=
m+V$ (Fig. 2d). One can say that effective momentum transfer $q_0=m_{\Lambda }-m$ 
and $q_0=V$ takes place. The nonzero momentum transfer is the common property of
the diagrams shown in Fig. 2. 

\begin{figure}[h]
%  \reflectbox{\includegraphics[height=.3\textheight]{golfer}}
  {\includegraphics[height=.25\textheight]{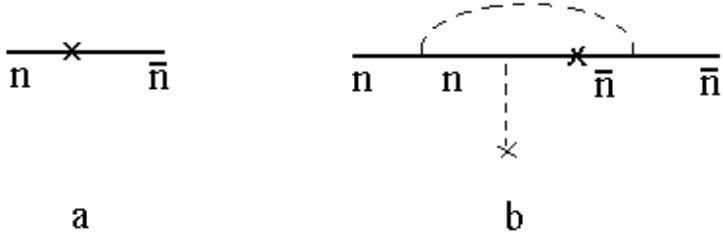}}
  \caption{Free-space $n\bar{n}$ transition ({\bf a}) and medium correction to
the $n\bar{n}$ transition vertex ({\bf b}).}
\end{figure}

The limit obtained for the free-space $n\bar{n}$ oscillation time is [10,14]
\begin{equation}
\tau _{n\bar{n}}>10^{16}\; {\rm yr}.
\end{equation}
In [10] and calculations made above the free-space $n\bar{n}$ transition operator has
been used. This is impulse approximation which is employed for nuclear $\beta $ decay,
for instance. The simplest medium correction to the vertex (or off-diagonal mass, or
transition mass) is shown in Fig. 3b. In this event the replacement should be made: 
\begin{equation}
\epsilon \rightarrow \epsilon _m=\epsilon (1+\Delta \epsilon ),
\end{equation}
$\Delta \epsilon  =\epsilon _{3b}/\epsilon $, where $\epsilon _{3b}$ is the correction
to $\epsilon $ produced by the diagram 3b. Then the limit becomes
\begin{equation}
\tau _{n\bar{n}}>(1+\Delta \epsilon )10^{16}\; {\rm yr}.
\end{equation}
Obviously, the $\Delta \epsilon $ cannot change the order of magnitude of $\tau 
_{n\bar{n}}$ since the $n\rightarrow \bar{n}$ operator is essentially zero-range one. 
The free-space $n\bar{n}$ transition comes from the exchange of Higgs bosons with the 
mass $m_H>10^5$ GeV [5]. Since $m_H\gg m_W$ ($m_W$ is the mass of $W$-boson), the
renormalization effects should not exceed those characteristic of nuclear $\beta $
decay which is less than 0.25 [15]. So the medium corrections to the vertex are 
inessential for us. 

We now turn to a discussion of interference between different processes. The diagram
2c cannot produce interference, since it contains $K$-meson in the final state.
Figure 2d is included in diagram 1a. As mentioned above, it was adduced only for
illustration of the effect on our results of possible fluctuations. For the rest of
the diagrams the significant interferences are unlikely because the final state in 
$\bar{n}N$ annihilation are very complicated configurations and persistent phase 
relations between different amplitudes cannot be expected. This qualitative picture
is confirmed by our calculations [16] for $\bar{p}$-nuclear annihilation. Moreover, 
in [16] the interference between the amplitudes which give the comparable 
contribution takes place. As shown later, the contribution of diagrams 2 is negligible.

The corrections given above should be compared with the results obtained for the
basic model. We adduce the main results corresponding to diagram 1a. (This problem
has been considered in [10]. Nevertheless, we should make additional comments since
the time-dependence of the processes shown in Figs. 1 and 2 is different (see below)
what is very unusual.) The interaction Hamiltonian is taken in the general form:
\begin{equation}
{\cal H}_I={\cal H}_{n\bar{n}}+{\cal H}.
\end{equation}
The notations are the same as in (2). Formally, in the lowest order in ${\cal H}_{n\bar{n}}$
the process amplitude $M_1$ is uniquely determined by the Hamiltonian (19):
\begin{equation}
M_1=\epsilon G_sM^{(n)}
\end{equation}
(see Fig. 1a), where the antineutron propagator $G_s$ is given by (9). So $M_1\sim 1/0$.
This is infrared divergence conditioned by zero momentum transfer in the $n\bar{n}$
transition vertex. For solving the problem the approach with a finite time interval (FTA) 
[17] is used. The problem is formulated on the interval $(t,0)$ which corresponds to the 
concrete conditions of the process under study. (Strictly speaking diagrams 2 should be 
calculated on the interval $(t,0)$ as well. However, for non-singular diagrams the FTA 
converts to $S$-matrix theory [10].) If
\begin{equation}
\Gamma t\gg 1
\end{equation}
then the process (1) probability $W_1(t)$ is 
\begin{equation}
W_1(t)\approx W_f(t)=\epsilon ^2t^2,
\end{equation}
where $W_f(t)$ is the free-space $n\bar{n}$ transition probability. Owing to the strong 
annihilation channel, $W_1(t)\approx W_f(t)$. The antineutron annihilates in a time $\tau 
_a\sim 1/\Gamma $. So condition (21) implies that the annihilation can be considered 
instantaneous: $\tau _a\sim 1/\Gamma \ll t$. 

Equation (22) corresponds to the limiting case (21). The opposite limiting case $\Gamma 
\rightarrow 0$ means that there is no annihilation and so $W_1=0$. This result follows 
from the equations of motion. In the intermediate range $\Gamma t\sim 1$ the distribution 
$W_1(t)$ is more {\em complicated} (see Eq. (64) and Sect. 8 of Ref. [10]). It can be 
expected that $W_1(t)<W_f(t)$, although no detailed calculations is possible.

The physical models used for diagrams 1 and 2 are identical. The sole (but fundamental) 
distinction of Fig. 1 with respect to the diagrams 2 is the zero momentum transfer in the 
first vertex. Also we would like to emphasize that result (22) is true for any neutron 
wave function since the zero momentum transfer in the $n\bar{n}$ transition vertex takes 
place in any case.

\section{Time-dependence and contribution of corrections}
Comparing (22) and (8) we have
\begin{equation}
\frac{W_{2a}(t)}{W_1(t)}=\frac{\Gamma }{m^2_{\pi }t},
\end{equation}
where we have put $m_{\Phi }=m_{\pi }$. Consequently, if
\begin{equation}
m^2_{\pi }t/\Gamma \gg 1,
\end{equation}
and $\Gamma t\gg 1$ (see (21)) then the contribution of diagram 2a is negligible.
(For the $n\bar{n}$ transition in nuclei conditions (21) and (24) are fulfilled since
in this case $\Gamma \sim 100$ MeV and $t=T_0=1.3$ yr, where $T_0$ is the observation
time in proton-decay type experiment [18].) In fact, it is suffice to hold condition
(21) only because it is more strong.

Similarly, with the help of (22) and (15) we conclude: if $V^2t/\Gamma \gg 1$ and 
$\Gamma t\gg 1$, the contribution of diagram 2d is negligible. So $W_{2a}\ll W_1$ and 
$W_{2d}\ll W_1$ because $W_{2a}\sim t$ and $W_{2d}\sim t$, whereas $W_1\sim t^2$. The 
quadratic time-dependence of the $W_1$ is due to the zero momentum transfer in the 
$n\bar{n}$ transition vertex. The heart of the problem is as follows.

For the processes with $q=0$ the $S$-matrix problem formulation $(\infty,-\infty)$
is physically incorrect [10]. This is apparent even from the limiting case ${\cal H}=0$:
if ${\cal H}_I={\cal H}_{n\bar{n}}$ (see (19)), the solution is periodic. It is obtained
by means of non-stationary equations of motion and not $S$-matrix theory. To reproduce
the limiting case ${\cal H}\rightarrow 0$, i.e. the periodic solution, we have to use
the FTA. Formally, the $S$-matrix approach is inapplicable because the perculiarity
$M_1\sim 1/0$ is unremovable one. The FTA is {\em infrared-free}.

If the problem is formulated on the interval $(t,0)$, the decay width $\Gamma $ cannot
be introduced since $\Gamma =\sum_{f\neq i}\mid S_{fi}(\infty,-\infty)\mid ^2/T_0$,
$T_0\rightarrow \infty $. This means that the standard calculation scheme should be
completely revised. (We would like to emphasize this fact.) The direct calculation by 
means of evolution operator gives the distribution (22).

The more physical explanation of the $t^2$-dependence is as follows. In the Hamiltonian
(2) corresponding to Fig. 2a we put ${\cal H}={\cal H}_{n\bar{n}}=0$. Then the virtual
decay $n\rightarrow n\Phi $ takes place. The first vertex of the diagram 2a dictates
the exponential decay law of the overall process shown in Fig. 2a. Similarly, in the
Hamiltonian (19) corresponding to Fig. 1a, we put ${\cal H}=0$. Then the free-space
$n\bar{n}$ transition takes place which is quadratic in time: $W_f(t)=\epsilon ^2t^2$. 
The first vertex determines the time-dependence of the whole process at least for small
$\Gamma $. (See, however, text below (22).) We also recall that even for proton decay
the possibility of non-exponential behavior is realistic [19-22].

In Fig. 2c the transition $n\rightarrow \bar{\Lambda }$ is depicted by 2-tail vertex.
Nevertheless, the diagram 2c is described by the exponential decay law because in the 
above-mentioned vertex the effective momentum transfer takes place. As a result, the 
structure of the amplitude $M_{2c}$ is the same as that of the $M_{2a}$ and $M_{2b}$. 
Indeed, let us compare (4) and (13). Put ${\bf p}={\bf q}=0$ for simplicity. Equation 
(5) becomes $G=1/(p_0-q_0-m+i0)$. On the other hand, if $q_0=m_{\Lambda }-m$, this
expression is the propagator of (13). Since the structure of propagators and amplitudes 
$M_{2a}$ and $M_{2c}$ is identical, the $M_{2c}$ corresponds to the effective decay. 
Due to the effective momentum transfer $q=(m_{\Lambda }-m,0)$ in the first vertex, the
diagram 2c is described by the exponential decay law as well.

Ratio (23) does not depend on $\epsilon $, whereas $W_{2b}/W_1$ contains two unknown
parameters $\epsilon $ and $\epsilon _{\Phi }$ and so the contribution of diagrams
1a and 2b cannot be compared numerically. Nevertheless, it is safe to argue that
for sufficiently large times $W_{2b}\ll W_1$ because $W_1(t)\sim t^2$, whereas
$W_{2b}(t)\sim t$. The same is true for Fig. 2c: $W_{2c}\ll W_1$.

Formally, the different time-dependence of the processes shown in Figs. 1 and 2 is
due to $q$-dependence of amplitudes. If $q$ decreases, the amplitudes $M_{2a}-M_{2d}$
increase; in the limit $q\rightarrow 0$ they are singular. The point $q=0$ corresponds 
to realistic process shown in Fig. 1a. The $t^2$-dependence of this process is the 
consequence of the zero momentum transfer. Due to this for sufficiently large times 
($t\gg 1/\Gamma $ for Fig. 2a, for example) the contribution of diagrams 2 is negligible. 
We also recall that the diagrams 2c and 2d give no contribution to the amplitude of the 
process under study. They have been considered for illustration of the role of additional 
baryon-number-violating processes (Fig. 2c) and possible fluctuations (Fig. 2d).

\section{Conclusion}
The diagrams shown in Fig. 2 correspond to the decays (Figs. 2a and 2b) or decay-type
processes (Figs. 2c and 2d). They are described by the exponential decay law. The
corresponding decay widths have been calculated. The probability of the process shown 
in Fig. 1a is quadratic in time: $W_1\sim t^2$ (see, however, text below (22)). This is 
due to the zero momentum transfer in the $n\bar{n}$ transition vertex. The point $q=0$ 
is the peculiar point of the $S$-matrix amplitude. If the amplitude is singular, the 
surprises can be expected. From this standpoint a departure from the exponential decay 
law comes as no surprise to us. It seems natural that for non-singular and singular 
diagrams the functional structure of the results is different, including $t$-dependence. 
The opposite situation would be strange.

To summarize, the diagrams 1a and 2 are fundamentally different. The time-dependence of 
the corresponding processes is different as well. Because of this for sufficiently large 
times the contribution of the diagrams 2 is negligible. The distribution (22) and limit 
(16) can be essentially changed only if there is no diagram 1a, which is out of question. 
For the $K^0\bar{K}^0$ oscillations in the medium the similar consideration can be done.

\end{document}